\documentclass{article}

\usepackage{arxiv}

\usepackage[utf8]{inputenc} 
\usepackage[T1]{fontenc}    
\usepackage{hyperref}       
\usepackage{url}            
\usepackage{booktabs}       
\usepackage{amsfonts}       
\usepackage{microtype}      
\usepackage{lipsum}

\title{Compressive analysis and the future of privacy}

\author{
  Suyash Shandilya \\
  Defence Research and Development Organisation\\
  New Delhi - 110054\\
  \texttt{suyash@ece.ism.ac.in} \\
}

\begin{document}
\maketitle

\begin{abstract}
\textit{Compressive analysis} is the name given to the family of techniques that map raw data to their smaller representation. Largely, this includes data compression, data encoding, data encryption, and hashing. In this paper, we analyse the prospects of such technologies in realising customisable individual privacy. We enlist the dire need to establish privacy preserving frameworks and policies and how can individuals achieve a trade-off between the comfort of an intuitive digital service ensemble and their privacy. We examine the current technologies being implemented, and suggest the crucial advantages of compressive analysis.
\end{abstract}


\section{Introduction}
As more and more technologies grow around models that require large amount of minute data, at some point we will need to step back and evaluate its repercussions on individual privacy. \emph{All of us are secure or none of us are} \texttt{ - Edward Snowden}. Even if a sufficiently large subset of the complex mesh of our social data is revealed, it compromises the privacy of even those who were never quite a part of it. We need to start developing frameworks of a system where there are limits to what an analytic software can and cannot know. And each individual should have a throttle over their own data. Simply ensuring secure and trustworthy systems won't solve the impending problem at large. In the following sections we re-evaluate the exigency of privacy preserving protocols. Key areas which are either booming or on the verge of a technological revolution, and also need special privacy concerns are then identified. Following that, current solutions to the issue are discussed. Towards the end we state the significance of compressive analysis as a holistic solution to the privacy problem.
\section{Pivotal trends and their need for privacy}
Comfort is the compass for consumer technology. This includes the convenience in instructing the system (parsing natural language), easier interface, bundled service suits, intuitive suggestions, personalised automations and suggestions, ad infinitum.
Additionally, comfort can also be defined as a peaceful state of mind. A state where a person is not agitated by prying advertisements, or obsequious suggestions. Where they can enjoy the repose of a private space, knowing well enough that their personal information is secure and entirely under their control - the leisure of \emph{privacy}.

Here we consider 4 major fields which will (continue to) grow at an accelerated pace in the coming times.
\subsection{Era of Data Analytics}
There is no denial to the advent of the data analytics industry over the past couple of decades, and this trend is far from slowing down. It helps in understanding user preferences, their perceptions, etc. More specifically, it's about quantitatively defining a user, whether it's an individual or a firm. Knowing their daily schedule, monitoring the quantity and nature of their daily subscription services like mail, contacts etc. In the personal domain, it even includes logging health data, finances, IoT device logs, etc. Typically, a user can benefit a lot from this data. They can improve their spending, health, work efficiency, and what not. The technology is now mature enough to predict your future 'footprints' in terms of decision and resource consumption, to a high degree of accuracy. Users can use these data to make more informed, statistically wiser decisions.
But this wisdom comes as a compromise of privacy. The user has to allow the services to access their raw data to benefit from the analysis. The service may assure non-distribution of access to other parties but that limits
\subsection{Human-Computer Interaction}
Human Computer Interaction (HCI) have always been the face of futuristic technologies. They cover all domains and aspects of how a (human) user interacts with a computer system. It includes common input devices like mouse,keyboard, to more advanced modules like motion capture, Brain Computer Interface (BCI) etc. They can be completely non-physical like an interactive AI-assistant. They all intend to make your interaction easier with the computer, and in the process learn fine intricacies of your commands - your voice, your way of speech, your physical gestures, your facial expressions, etc. Their efficacy as an interface hinges on the data they retain about you, and their ability to extract relevant features. Given the fact that these human actions undergo subtle changes with time (and across users in a multi-user scenario), it is difficult to request it to stop logging data without compromising efficiency. This makes HCI evolution particularly poised in privacy concerns. Huge logs of raw BCI data mean that the computer would know more objectively, the thought-action sequence of a user than the user themselves.
Unfortunately there has been very limited research concerning privacy preserving HCI methods.
\subsection{Emotional Intelligence}

As on date, we have an outlandish amount of ever increasing computation power. Any feasible problem which can have a procedural solution can be solved. For most of our common problems, the IQ of our current systems, is pretty satisfactory. Construing human emotions however, has been a pretty elusive task. To break into the rooms of every bloke in the world, our systems will need to connect with us at a less logical and more emotional level. The primary bottleneck has been rigorous definitions of emotional boundaries. The scientific understanding, detection and classification of 'emotion' or 'emotional response' has largely been based on physiological signals like ECG \cite{emoecg}, EEG \cite{emoeeg}, breathing rate \cite{eqradio}, etc. Typically one couldn't access these signals in a mundane scenario without bundling the subject with multiple probes. However, it is demonstrated in \cite{eqradio} that such inferences are possible wirelessly, and provide remarkable accuracy. Wireless sensing \cite{wilesssense} has enabled a completely unintrusive sensing scheme which also leads to less errors as the subjects don't feel the physical agitation and mental sense of being under observation. While beneficial from a health perspective, such technology is a serious threat to privacy. It can be very conveniently deployed to read any unaware individual's precise breathing rate and pulse rate counts.
Although less objective than physiological signals, Humans normally don't construe emotions by reading each other's breathing rate or heartbeats. We simply rely on our empirical understanding of facial expressions. But implementing such a system in public domain would be a paragon of privacy violation. Nevertheless emotional intelligence is an imperative course of future. The society still has a long way to go towards mainstreaming mental health issues, but things are on a favourable course. There are many projects aimed at examining a subject's emotions and providing tailored therapeutic treatment. 
\cite{ppfacerec} addresses the privacy problem using a multicomputing model. \cite{eqcrowd} makes an attempt at reading emotions via a system of individual participation. Other solutions of the problem are discussed in section \ref{sec:ca}

\subsection{Internet of Things}
The Internet of Things (IoT) is another major technology that has already started manifesting in our lives. It consists of all the hardware and software in an individual's ambience collaborating towards an optimised routine. Despite being industrialised at a fast pace, it still remains as a popular spot for not just privacy violations, but security violations as well \cite{mirai}. The shear number of these primitive yet garrulous IoT devices make security guarantees difficult to enforce. Nevertheless, there have been remarkable developments in the field of IoT security and privacy \cite{secureiot}.

Notwithstanding the many roadblocks in achieving privacy preservation, \textit{trust} in a \textbf{secure} system can ensure that the data being collected by the system is not being shared unsolicitedly with third parties for personal profit. In fact, there are good reasons to secure centralised data collection like healthcare, policy planning. Additionally, most of the 'intelligent' system can not afford to run locally (i.e. as completely standalone, non-communicating machines). The manufacturers would require data from vast sources to develop more robustly responsive systems. It is seemingly impossible to develop better systems without regular due feedback from our existing ones.
Having said that, centralised data collection by any organisation - no matter how secure or trustworthy - must be impeded. There are many fictional dystopian descriptions of societies where certain organisations controlled micromovements of their subjects by the virtue of large scale data access. Although exaggerated and dramaitsed, these scenarios provide reference marks for policy makers and technology designers to consider while developing modern systems.

\section{State of the art solutions} \label{sec:homo}
\textbf{Security} is the parent condition for privacy. There can be no question about privacy if there is no security.

Before discussing the current solutions, one should be acquainted with the idea of \textbf{differential privacy}. It is a privacy guarantee which requires that computations be formally indistinguishable when run with and without any one record,almost as if each participant had opted out of the data set \cite{dp}. They discuss its implementation by adding a calculated amount of random noise, either to the input data, or to the inference accessed from the database \cite{dpnoise}.\\
The idea of \textbf{homomorphic encryption} is to "preserve the shape" of the plain text in the cipher text. This means that the ciphertext can be operated (with limited operators like addition and multiplication \cite{homothesis}) to form inferences, without the need for decryption. This property makes it perfect for privacy preserving applications and has rightly found many applications in e-voting \cite{homoelec},biometric-medical data processing \cite{ppfacerec,ppecg}, recommender systems \cite{pphomoreco,ppreco} and in multicomputation systems.\cite{pphomomulticompute,eqcrowd}. Interestingly, it is a lattice-based cryptography technique \cite{lattice} which is among the forerunners for post quantum cryptography algorithms \cite{pqc}. Thus along with privacy preservation, it is well suited as an advanced security solution as well.
A similar approach to securing data is conceptualised in \textbf{cryptDB} \cite{cryptdb}. In it, the authors have devised a system where along with the database, the requests and results are encrypted as well.
\textbf{Decentralisation} is about as globally accepted as a solution, as the privacy preservation problem itself. By far the most popular implementation of decentralisation is \textbf{cryptocurrency}. What had been centralised for centuries, was conceived and implemented as a decentralised, abstract entity. This has often been purported as such an extreme solution to privacy that it disturbed the authorities as another serious problem of lack of control.
Despite being notoriously data mongering of all modern deep learning models, neural networks are still a rage in the industry. Their incomparable dexterity in establishing landmarks in the field of machine learning has assured that their trend won't recede anytime soon. Nevertheless, their large appetite for data raises serious privacy concerns. \textbf{Split learning} \cite{split} and \textbf{Federated learning} \cite{fed} have emerged as the most popular solution to privacy preserving learning. They both achieve this by implementing standard learning in decentralised models. \cite{splitfedcomp} provides an excellent comparison between the two models of learning.
Despite all the technical support the industry might extend, privacy remains fundamentally a \textbf{policy} level problem.  Digital security solutions are not an immediate concern in our time as the current systems have been working satisfactorily well and any breach can be quickly located and patched. But if a secure system is asked to share the information it houses - through proper channel and authentication - it will comply. It is up to the service agreement between the individual and its service provider to christen what data will be shared with whom and when. Unfortunately, due to lack of awareness, people don't heed the service agreement much and simply consent to what is being offered. Often there aren't many choices provided by the service provider. And even for a genuinely aware user, it is not easy to parse the legalese of each agreement they sign with each of their service provider. Projects like the Usable Privacy project \cite{upp} help the users to parse their agreements more conveniently but stronger policy level enforcements will still be required if the problem is to be addressed globally.

\section{Compressive analysis}

Compressive analysis is a way to draw relevant inferences from a compressed data, without decompressing it. The compressed data is also encrypted in some way. Typically, neither decryption nor decompression of any kind is expected, or even made possible. This means that a lot of data is already lost in the process of storing or acquisition (more on this later). Ideally, one can expect that only the parts of data that are required for drawing relevant inferences are retained in the compression.
This can be achieved by training the compression algorithm on large amounts of dummy data which contains the raw information along with the sensitive labels. We can then draw simultaneous analyses and private information out of the compressed data. They will work as adversial networks where the analyser function will try to minimise the number of components required, while the devious function will attempt at extracting private information from the set components. Additionally, it gives a throttle to the user to determine their own degree of privacy against different analyser softwares.
It is also particularly useful where \textbf{expurgation} can not be a privacy solution. The challenge will be to guarantee \textit{differential privacy} against auxillary information which is available to the attacker via ulterior sources \cite{dp}.

\subsection{Hash analysis}
Hashing is a tool which takes in a large chunk of data and returns a small, finite bit long 'digest' of the data. The function is one-sided thus it is impossible to reconstruct the original data from the hash. It is designed to be incoherent enough that different data have different hash and hash collisions are extremely rare. The most commonly known hashes like SHA-256, SHA3 family of hashes, are all cryptographic hashes. They work on avalanche effect which means, any bit changed results in large change in the digest of the data. \textbf{Perceptual hashes} \cite{phash} however are commonly used in industry to classify similar (same but transformed in some way) images together. They are used to test duplicacy of copyright images, and often for fast image searches as well.
These perceptual hashes can also be used for specific feature detection when the original data source does not want its data to be seen. However this method requires that the data is read during the hashing process. Following is a solution which ensures that the data is never stored digitally starting from the acquisition stage.

\subsection{Compressive Aquisition} \label{sec:ca}

\emph{Compressive Sensing} is a relatively young technology which posits that accurate reconstrution (or erroneous reconstruction where the error is limited by the level of noise in sampling) is possible via random linear compressed samples \cite{cs}. Interested readers may refer to a well summarised yet basic introduction \cite{cstut}. This reconstitutes compression as an embedded acquisition step instead of an impending post-processing.
It has found some applications in the privacy preservation domain as well. One of the earliest works I encountered was \cite{pprs} where the authors use compressed sensing as a matrix masking \cite{matrixmasking} technique and prove theoretically that regression analysis can be deployed on a transformed dataset as effectively as in its raw form. In a sense, it is similar to homomorphic transformation. Related works can be found in  \cite{ppra,ppra2}.
The theoretical foundations of compressive sensing were first laid out by Donoho \cite{donoho}, Candes and Tao \cite{candromtao}, but the most popular practical demonstration was shown by Duarte, et al. in \cite{spc} - A single pixel camera which acquired few random compressed samples to reconstruct images. In \cite{smash}, the concept of \emph{smashed filters} is introduced which shows that 'target classification' can be achieved using a maximum likelihood estimator on the compressed samples themselves. These ideas and subsequent demonstration give rise to the pertinent concept of \textbf{compressive acquisition}.

If the process of acquisition itself can be designed to be random and lossy, while allowing for reasonable reconstruction nevertheless, much of the sensitive data will never even be digitised, let alone thieved. The first privacy oriented demonstration of this concept was on print error detection in \cite{mima}. It is shown that even when the degree of compression is less than 1\%, very high accuracy classification can be achieved. Further, even if the decryption key (sensing matrix, in this case) is known and is used to reconstruct the image via the best possible algorithm, the result is practically unusable.

Such techniques can in general be deployed in various systems. For example, a set of monte carlo linear measurements (apropos to compressive sensing) from a wireless sensor network can be used instead of continuous values to ensure privacy preservation. Thus, among all the possible solutions discussed in this section, compressive acquisition would be the most promising privacy preserving approach as it eliminates the existence of private information in a dataset altogether.

\section{Conclusion}
We began with acknowldedging the dire need of privacy in current times. We noted four major fields that are going to grow at an accelerated pace and will have serious privacy concerns namely: Data analytics, Human Computer Interaction, Emotion recognition and Internet of Things. We then looked at some of the common approaches that are currently being deployed to combat the privacy problem and noted that most of them are centred around security and decentralisation. We then moved to indtroduce compressive analysis which is a family of all the different kinds of approach which attempt to remove all the privacy sensitive infomration from a dataset, before storing it for analytics. Among it were hashes, and my novel idea of compressive acquisition which is based on the principle of compressive sensing and it aims to acquire a data in a randomised, linearly compressed manner so that no meaningful reconstruction is possible while guaranteeing satisfactory analytical result. We concluded that compressive acquisition, whenever applicable, is the most effective solution of assuring privacy.

\bibliographystyle{unsrt}  
\bibliography{references}

\begin{thebibliography}{10}

\bibitem{emoecg}
F.~{Agrafioti}, D.~{Hatzinakos}, and A.~K. {Anderson}.
\newblock Ecg pattern analysis for emotion detection.
\newblock {\em IEEE Transactions on Affective Computing}, 3(1):102--115, Jan
  2012.

\bibitem{emoeeg}
Panagiotis~C Petrantonakis and Leontios~J Hadjileontiadis.
\newblock Emotion recognition from eeg using higher order crossings.
\newblock {\em IEEE Transactions on Information Technology in Biomedicine},
  14(2):186--197, 2009.

\bibitem{eqradio}
Mingmin Zhao, Fadel Adib, and Dina Katabi.
\newblock Emotion recognition using wireless signals.
\newblock In {\em Proceedings of the 22nd Annual International Conference on
  Mobile Computing and Networking}, pages 95--108. ACM, 2016.

\bibitem{wilesssense}
Axel Scherer, Muhammad Mujeeb-U-Rahman, Meisam~Honavar Nazari, and
  Muhammad~Musab Jilani.
\newblock Minimally invasive wireless sensing devices and methods, January~8
  2019.
\newblock US Patent 10,172,520.

\bibitem{ppfacerec}
Zekeriya Erkin, Martin Franz, Jorge Guajardo, Stefan Katzenbeisser, Inald
  Lagendijk, and Tomas Toft.
\newblock Privacy-preserving face recognition.
\newblock In Ian Goldberg and Mikhail~J. Atallah, editors, {\em Privacy
  Enhancing Technologies}, pages 235--253, Berlin, Heidelberg, 2009. Springer
  Berlin Heidelberg.

\bibitem{eqcrowd}
Zeki Erkin, Jie Li, Arnold~POS Vermeeren, and Huib de~Ridder.
\newblock Privacy-preserving emotion detection for crowd management.
\newblock In {\em International Conference on Active Media Technology}, pages
  359--370. Springer, 2014.

\bibitem{mirai}
Manos Antonakakis, Tim April, Michael Bailey, Matt Bernhard, Elie Bursztein,
  Jaime Cochran, Zakir Durumeric, J~Alex Halderman, Luca Invernizzi, Michalis
  Kallitsis, et~al.
\newblock Understanding the mirai botnet.
\newblock In {\em 26th $\{$USENIX$\}$ Security Symposium ($\{$USENIX$\}$
  Security 17)}, pages 1093--1110, 2017.

\bibitem{secureiot}
Kevin Kiningham, Mark Horowitz, Philip Levis, and Dan Boneh.
\newblock Cesel: Securing a mote for 20 years.
\newblock In {\em Proceedings of the 2016 International Conference on Embedded
  Wireless Systems and Networks}, EWSN '16, pages 307--312, USA, 2016. Junction
  Publishing.

\bibitem{dp}
Dwork Cynthia.
\newblock Differential privacy.
\newblock {\em Automata, languages and programming}, pages 1--12, 2006.

\bibitem{dpnoise}
Cynthia Dwork, Frank McSherry, Kobbi Nissim, and Adam Smith.
\newblock Calibrating noise to sensitivity in private data analysis.
\newblock In Shai Halevi and Tal Rabin, editors, {\em Theory of Cryptography},
  pages 265--284, Berlin, Heidelberg, 2006. Springer Berlin Heidelberg.

\bibitem{homothesis}
Craig Gentry.
\newblock {\em A fully homomorphic encryption scheme}.
\newblock PhD thesis, Stanford University, 2009.
\newblock \url{crypto.stanford.edu/craig}.

\bibitem{homoelec}
Martin Hirt.
\newblock Receipt-free k-out-of-l voting based on elgamal encryption.
\newblock In {\em Towards Trustworthy Elections}, pages 64--82. Springer, 2010.

\bibitem{ppecg}
Mauro Barni, Pierluigi Failla, Riccardo Lazzeretti, Ahmad-Reza Sadeghi, and
  Thomas Schneider.
\newblock Privacy-preserving ecg classification with branching programs and
  neural networks.
\newblock {\em IEEE Transactions on Information Forensics and Security},
  6(2):452--468, 2011.

\bibitem{pphomoreco}
Z.~{Erkin}, T.~{Veugen}, T.~{Toft}, and R.~L. {Lagendijk}.
\newblock Generating private recommendations efficiently using homomorphic
  encryption and data packing.
\newblock {\em IEEE Transactions on Information Forensics and Security},
  7(3):1053--1066, June 2012.

\bibitem{ppreco}
Arjan J.~P. Jeckmans, Michael Beye, Zekeriya Erkin, Pieter Hartel, Reginald~L.
  Lagendijk, and Qiang Tang.
\newblock {\em Privacy in Recommender Systems}, pages 263--281.
\newblock Springer London, London, 2013.

\bibitem{pphomomulticompute}
R.~L. {Lagendijk}, Z.~{Erkin}, and M.~{Barni}.
\newblock Encrypted signal processing for privacy protection: Conveying the
  utility of homomorphic encryption and multiparty computation.
\newblock {\em IEEE Signal Processing Magazine}, 30(1):82--105, Jan 2013.

\bibitem{lattice}
Oded Regev.
\newblock Lattice-based cryptography.
\newblock In {\em Annual International Cryptology Conference}, pages 131--141.
  Springer, 2006.

\bibitem{pqc}
Daniel~J Bernstein.
\newblock Introduction to post-quantum cryptography.
\newblock In {\em Post-quantum cryptography}, pages 1--14. Springer, 2009.

\bibitem{cryptdb}
Raluca~Ada Popa, Catherine Redfield, Nickolai Zeldovich, and Hari Balakrishnan.
\newblock Cryptdb: protecting confidentiality with encrypted query processing.
\newblock In {\em Proceedings of the Twenty-Third ACM Symposium on Operating
  Systems Principles}, pages 85--100. ACM, 2011.

\bibitem{split}
Praneeth Vepakomma, Otkrist Gupta, Tristan Swedish, and Ramesh Raskar.
\newblock Split learning for health: Distributed deep learning without sharing
  raw patient data.
\newblock {\em arXiv preprint arXiv:1812.00564}, 2018.

\bibitem{fed}
Jakub Kone{\v{c}}n{\`y}, H~Brendan McMahan, Felix~X Yu, Peter Richt{\'a}rik,
  Ananda~Theertha Suresh, and Dave Bacon.
\newblock Federated learning: Strategies for improving communication
  efficiency.
\newblock {\em arXiv preprint arXiv:1610.05492}, 2016.

\bibitem{splitfedcomp}
Abhishek Singh, Praneeth Vepakomma, Otkrist Gupta, and Ramesh Raskar.
\newblock Detailed comparison of communication efficiency of split learning and
  federated learning, 2019.

\bibitem{upp}
Abhilasha Ravichander, Alan~W Black, Shomir Wilson, Thomas Norton, and Norman
  Sadeh.
\newblock Question answering for privacy policies: Combining computational and
  legal perspectives.
\newblock In {\em Proceedings of the 2019 Conference on Empirical Methods in
  Natural Language Processing and the 9th International Joint Conference on
  Natural Language Processing (EMNLP-IJCNLP)}, pages 4946--4957, Hong Kong,
  China, November 2019. Association for Computational Linguistics.

\bibitem{phash}
Xia-mu Niu and Yu-hua Jiao.
\newblock An overview of perceptual hashing.
\newblock {\em Acta Electronica Sinica}, 36(7):1405--1411, 2008.

\bibitem{cs}
Richard~G Baraniuk.
\newblock Compressive sensing.
\newblock {\em IEEE signal processing magazine}, 24(4), 2007.

\bibitem{cstut}
E.~J. {Candes} and M.~B. {Wakin}.
\newblock An introduction to compressive sampling.
\newblock {\em IEEE Signal Processing Magazine}, 25(2):21--30, March 2008.

\bibitem{pprs}
Shuheng Zhou, John Lafferty, and Larry Wasserman.
\newblock Compressed and privacy-sensitive sparse regression.
\newblock {\em IEEE Transactions on Information Theory}, 55(2):846--866, 2009.

\bibitem{matrixmasking}
George~T Duncan, Robert~W Pearson, et~al.
\newblock Enhancing access to microdata while protecting confidentiality:
  Prospects for the future.
\newblock {\em Statistical Science}, 6(3):219--232, 1991.

\bibitem{ppra}
C.~{Wang}, B.~{Zhang}, K.~{Ren}, and J.~M. {Roveda}.
\newblock Privacy-assured outsourcing of image reconstruction service in cloud.
\newblock {\em IEEE Transactions on Emerging Topics in Computing},
  1(1):166--177, June 2013.

\bibitem{ppra2}
Cai Chen, Manyuan Zhang, Huanzhi Zhang, Zhenyun Huang, and Yong Li.
\newblock Privacy-preserving sensory data recovery.
\newblock In {\em 2018 17th IEEE International Conference On Trust, Security
  And Privacy In Computing And Communications/12th IEEE International
  Conference On Big Data Science And Engineering (TrustCom/BigDataSE)}, pages
  1646--1650. IEEE, 2018.

\bibitem{donoho}
D.~L. {Donoho}.
\newblock Compressed sensing.
\newblock {\em IEEE Transactions on Information Theory}, 52(4):1289--1306,
  April 2006.

\bibitem{candromtao}
E.~J. {Candes}, J.~{Romberg}, and T.~{Tao}.
\newblock Robust uncertainty principles: exact signal reconstruction from
  highly incomplete frequency information.
\newblock {\em IEEE Transactions on Information Theory}, 52(2):489--509, Feb
  2006.

\bibitem{spc}
Marco~F Duarte, Mark~A Davenport, Dharmpal Takhar, Jason~N Laska, Ting Sun,
  Kevin~F Kelly, and Richard~G Baraniuk.
\newblock Single-pixel imaging via compressive sampling.
\newblock {\em IEEE signal processing magazine}, 25(2):83--91, 2008.

\bibitem{smash}
Mark A.~Davenport, Marco Duarte, Michael B.~Wakin, Jason N.~Laska, Dharmpal
  Takhar, Kevin Kelly, and Richard Baraniuk.
\newblock The smashed filter for compressive classification and target
  recognition - art. no. 64980h.
\newblock {\em Proceedings of SPIE}, 6498, 02 2007.

\bibitem{mima}
Suyash Shandilya.
\newblock Minimizing acquisition maximizing inference - a demonstration on
  print error detection.
\newblock Springer, 2019.

\end{thebibliography}
\end{document}